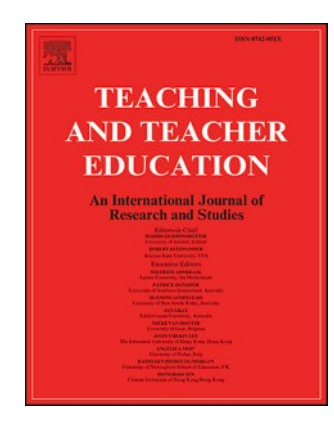

Research paper

# Responsive professional development: A facilitation approach for teachers' development in a physics teaching community of practice

Hamideh Talafian *, Morten Lundsgaard, Maggie Mahmood, Devyn Shafer, Tim Stelzer, Eric Kuo

*University of Illinois Urbana-Champaign, Department of Physics, 1110 W Green St, Urbana, IL, 61801, USA*

A R T I C L E I N F O

A B S T R A C T

Creating learning environments that can accommodate teachers' diverse needs is challenging because responsive elements are not clearly defined or identified. This study identified responsive teacher professional development (PD) elements by taking a phenomenological approach. Using surveys and interviews with 13 high school physics teachers in a PD program at a Midwestern university, we identified responsive features such as practicality, flexibility, and accessibility core to the enactment of a responsive PD. Other features were opportunities for community engagement, pedagogical support, and professional growth, which were aligned with the benefits of engagement in a Community of Practice model incorporated in this work.

## 1. Introduction

There is a worldwide consensus among researchers and practitioners that professional development (PD) for teachers is an effective means of improving their classroom instruction and student achievement (Ball & Cohen, 1999; Darling-Hammond & McLaughlin, 1995; Skilbeck & Connell, 2003; Tripp, 2004). In service of this goal, many teacher PD programs in the U.S. focus on dissemination and training with new instructional tools and pedagogies yet may not adequately consider teachers' varied needs across different career stages (Coppe et al., 2024; Darling-Hammond et al., 2010). This approach risks disconnecting PD efforts from teachers' knowledge, beliefs, or local realities of their classroom context. This is one reason why research on the effectiveness of teacher PD has warned against one-shot, short, highly structured PD programs (Desimone, 2009; Yoon et al., 2007). By contrast, teacher PD can attend to teachers' diverse needs in various contexts to support rich learning experiences in PD settings. In this perspective, high-quality PD is not defined by a single characteristic but rather by a combination of features aligned with teachers' needs and interests (Ehrenfeld, 2022; Opfer & Pedder, 2011) that can contribute to the eventual success of the program.

The Communities of Practice (CoP) framework introduced by Lave and Wenger (1991) represents one approach with the potential to embrace diverse levels of experience and expertise in teacher learning in PD settings. In teaching CoPs, teacher learning is theorized to happen in an apprenticeship manner where more experienced teachers are placed in mentoring groups to share their experiences with novice teachers. While this structure offers novices the opportunity to move from peripheral to central roles within the community (Lave & Wenger, 1991), it often overlooks veteran's learning, creates an imbalanced power hierarchy, and can even hinder learning in such communities (Eschar-Netz & Vedder-Weiss, 2020; Sutton & Shouse, 2018). Additionally, the CoP framework, in practice, may fail to ensure the inclusion of diverse perspectives from teachers with varying backgrounds and experience levels. Hence, in this work, we present a responsive model of PD facilitation in a physics teaching CoP that aims to overcome these limitations of traditional conceptions of teacher CoP.

We conceptualize responsiveness in a PD setting as a facilitation strategy that is both *attentive* and *adaptive* in support of teachers' diverse backgrounds, knowledge, interests, affect, and needs. We remained attentive to teachers' needs by creating multiple opportunities for teachers to communicate their needs and adaptive by tailoring PD activities to reflect these needs. These adaptations occurred at the micro (e.g., regrouping teachers based on daily survey results) and macro levels (e.g., co-designing the structure of in-person PD workshops). We hypothesize that this responsive PD program, where PD coordinators co-design and co-facilitate PD activities with teachers, can better incorporate the perspectives and needs of teachers with varying backgrounds, experiences, and needs.

Given that teachers are central to this process, we sought their perspectives by posing the following research question: What do physics teachers perceive as the salient features of a PD program taking a responsive approach to fostering a teacher community of practice? Unlike other studies that assess the effectiveness of predefined strategies in teacher PD, our approach allowed teachers to identify and recognize PD features that matter most to them.

* Corresponding author.
*E-mail address:* talafian@illinois.edu (H. Talafian).

We believe taking a responsive approach with both attentive and adaptive elements empowers both novice and experienced teachers to actively contribute to PD design and facilitation, thereby fostering more productive and engaging peer interactions within the CoP. Additionally, this model can meet the diverse needs of educators, allowing for more personalized learning experiences that are directly relevant to their specific teaching contexts.

## 2. Background and theory

Designing PD for teachers with diverse academic backgrounds and classroom experiences can be a daunting task. Success in this endeavor requires attending to multiple aspects of teacher learning at both the individual and community levels. First, we describe how our PD has been informed by previous works within teacher learning and PD development. Then we situate our study within the literature on CoP and responsive teaching.

### *2.1. Teacher learning and professional development: linear and holistic views*

Previous models of teacher PD can be categorized into two main groups: linear and holistic models. The linear models, conceptualized as process-product or causality models by Opfer and Pedder (2011), view teacher learning as the mediator that connects PD to student learning. Under this view, the success of a PD program is measured by its ability to change teacher practice and, subsequently, enhance student learning. This linear conceptualization of teacher learning has been used to investigate anticipated student learning outcomes based on the characteristics of PD, such as the content, duration, or facilitation details of the PD program (Ball & Cohen, 1999; Carpenter et al., 1989; Desimone, 2009; Garet et al., 2001; Goldenberg & Gallimore, 1991; Yoon et al., 2007).

On the other hand, holistic models draw our attention to the complexities of teacher learning in PD settings and beyond. Ehrenfeld (2022) argues that linear models oversimplify the interconnected, complex web of teacher learning experiences. These complexities have been referred to as "nested" structures that involve systems within systems (Opfer & Pedder, 2011; Stollar et al., 2006). These more holistic studies highlight the complex contributing factors that lead to success or failure in teachers' implementations of PD practices. Deviating from the linear causal models of teacher learning, holistic models view teacher learning as a cyclic process of PD and implementation, where change happens along the way and takes time (Clarke & Hollingsworth, 2002; Clarke & Peter, 1993). This holistic approach has been employed to understand the complexities of teacher learning in professional learning communities aligned with the CoP framework (Blanton & Stylianou, 2009; Cochran-Smith & Lytle, 1999; MacPhail et al., 2014; McLaughlin & Talbert, 1993; Thomas et al., 1998). Teachers can recognize linearly conceived PD experiences as disconnected and distinct from their own professional learning because they do not consider teachers' sense of purpose, sense of responsibility, and personal fulfilment (Taylor, 2023). In one study of teachers' self-reported perceptions of PD impacts, a lack of contextual fit between PD pedagogies and teachers' classroom contexts and a lack of respect for teacher agency were identified as barriers to accepting and applying PD pedagogies (McChesney & Aldridge, 2021).

### *2.2. Teacher learning in professional communities of practice*

Looking at teacher PD through the situative lenses of the CoP framework allows us to see the complexities of learning as a social process (Lave & Wenger, 1991; Wenger, 1998). In this model, learning has been characterized in an apprenticeship fashion between old-timers and newcomers in preparation for the newcomer's eventual transition to taking a core role in the community (Lave & Wenger, 1991). The participants come together over a shared value or subject of interest (domain or joint enterprise) and form social interaction and relationships (community) by sharing knowledge and developing a shared repertoire including tools, documents, ideas, etc. (practice) (Wenger, 1998). Over time the newcomers' learning changes to result in what has been characterized as shifts in identity (Holland et al., 1998; Lave & Wenger, 1991). Opfer and Pedder (2013) argue that the CoP model can capture the complexities of teacher learning in PD between teachers who share the same enterprise.

From the PD facilitation perspective, designing meaningful learning experiences in a CoP should involve engaging learners as active participants rather than passive recipients of knowledge (Moje, 2015). This could be translated as seeing teachers as individuals responsible for their own understanding (Putnam & Borko, 1997). In such environments, the community becomes a space for sharing problems, strategies, and personal stories; hence, learning becomes a collective experience (Opfer & Pedder, 2013). Promoting challenging but safe spaces for sharing among teachers has been shown to increase teachers' risk-taking, their positive perceptions of learning in communities, and their willingness to try new practices (Patton & Parker, 2017; Parker et al., 2010; Deglau & O'Sullivan, 2006). Other strategies to meaningfully engage teachers in a CoP include mentoring (Patton et al., 2005), developing a common language (Bowes & Tinning, 2015; Gast et al., 2017), emphasizing teacher reflection (Pareja Roblin & Margalef, 2013) and establishing action research (Carli & Pantano, 2023; Norton et al., 2011) around a shared value. In physics teaching, such communities have proven effective for developing "productive habits" (Etkina et al., 2017) and lessening physics teacher attrition (Etkina, 2015).

Even by incorporating such strategies that actively engage teachers in their own learning, designing meaningful professional experiences for teachers using the CoP framework presents challenges. There may be tensions in maintaining diverse perspectives in the community. In response and to avoid such tensions, pseudo-communities can take shape, wherein the participants show excessive agreement, which can create an illusion of consensus (Grossman et al., 2001). Additionally, Lave and Wenger's original conceptualization of CoP (1991) focuses on how novices learn from experienced veterans. Designing from this view of CoP can neglect veteran's learning, create an imbalanced power hierarchy, and in some instances, threaten teacher learning in such communities (Eshchar-Netz & Vedder-Weiss, 2021; Sutton & Shouse, 2019). In response, some research has turned to focus on how experienced teacher leaders in physics teacher learning communities can benefit from collaborating with novices and can develop professionally through their leadership roles (Carli & Pantano, 2023; Levy et al., 2021). In our research and practice, we acknowledge these tensions and take a responsive approach to developing a teacher community—one that involves teachers in the process of co-design and brings their interests to the core.

### *2.3. Responsive professional development: expanding conceptualizations*

Responsive professional development has roots in "responsive teaching" and was first introduced in K-12 teaching (Robertson, Scherr, & Hammer, 2015; Thompson et al., 2016) in support of attending to students' thinking to engage them in disciplinary reasoning (Coffey, Hammer, Levin, & Grant, 2011; Hammer et al., 2012; Thompson et al., 2016). Responsive teaching is a cognitive-oriented pedagogical approach in which instructors build upon learners' reasoning to tailor their instruction (Ball, 1993; Robertson et al., 2015; Sherin & van Es, 2005). In this approach, the instructor not only attends to the learners' disciplinary ideas but also attempts to understand the events from the learners' perspective (Robertson et al., 2015). Responsiveness was later adopted by teacher education programs as a facilitation strategy for fostering teacher learning (Maskiewicz, 2015; Watkins et al., 2017, 2020). Yet, the definition of responsive PD programs is broader. Such definitions may range from applying cognitive strategies of responsive teaching to taking an adaptive strategy in PD design and facilitation. Hence, it is important to define in what sense we deem our program to be responsive.

Here, we characterize our *Responsive Professional Development* (RPD) approach as an adaptive facilitation approach that bases PD activities on the needs, interests, affects, and backgrounds of teachers, who each have specialized knowledge of their particular teaching context. This facilitation approach is aligned with design-based or adaptive PD models (Penuel et al., 2011; Trautmann & MaKinster, 2010). Previous research has shown that adaptive PD models that cater to teachers' perceived needs are effective (Coppe et al., 2024; Gabriel et al., 2011; Joubert & Sutherland, 2008; Petrie & McGee, 2012). Yet, just as research has critiqued how the traditional CoP model overfocuses on novice development, research on adaptive PD also cautions against overfocusing on novice teachers' needs and highlights the importance of including mid- and late-career teachers' needs as well (Bressman et al., 2018; Day, 2012).

The RPD approach not only attends to the complexities of teacher learning within PD settings but also to the critiques of the theoretical shortcomings of the CoP framework. The goal of the RPD approach is to minimize learning barriers resulting from differing perspectives by catering to the diverse needs of teachers at all career stages, with varied academic backgrounds and years of experience. Therefore, in this work, we seek teachers' perceptions of RPD elements and how these elements attended to their needs, interests, affects, and backgrounds.

### 2.4. Description of the partnership program

The Illinois Physics and Secondary Schools (IPaSS) program is a partnership between the University of Illinois and high school physics teachers in the state of Illinois. A core goal of the program is to make science, technology, engineering, and mathematics (STEM) fields more accessible to students by equipping high school teachers with high-quality instructional resources and peer support for teaching physics. In service of this goal, the curricular resources and PD sessions go hand in hand to support teachers in their diverse contexts and needs.

The *curricular component* of the program gives teachers free access to university-level physics materials including an SmartIllinois and iOLab device currently deployed in teaching introductory physics courses in the partnering university. These resources cover all mechanics, electricity, and magnetism appropriate for high school-level physics and are all research-based. However, the teachers are not obliged to implement materials made available through the program. In addition, a major part of the PD activities involves teachers sharing their pedagogies and materials with each other. The PD facilitators create opportunities for teachers to share a range of content and pedagogical approaches during PD meetings, focusing on highlighting novel, student-centered teaching approaches. Many pedagogical strategies and materials that teachers share, though not all, are direct products of or are inspired by educational research. Some examples of research-backed pedagogies shared by teachers are inquiry-based instruction (McDermott et al., 1996), the Investigative Science Learning Environment (ISLE) (Etkina & Van Heuvelen, 2007), and flipped-style instruction (Prasetyo et al., 2018). Appendix A shows an example of an in-person program schedule with the content and pedagogies introduced and discussed over one day of the program.

The *PD component* of the program connects teachers to a wide network of peers and university support via a blended virtual and in-person format. In the program, teachers complete roughly a total of 100 h of PD meetings per year for four years. The program's yearly PD cycle starts with three events during the summer: two four-day long, in-person summer workshops in June and August, respectively, and a set of three 2-h online meetings between the June and August sessions. Regular PD sessions throughout the year support teachers' in-class implementation of materials and pedagogical strategies introduced during summer PD (Fig. 1). Online meetings take place throughout the school year to provide teachers with continued support that started in the summer. Over the past four years of the program, the number of teachers involved has grown from four to 40, and the meetings have evolved to accommodate teachers' diverse curricular needs and incorporate their suggestions for facilitation. Each year, a new cohort of teachers joins the returning group, allowing cross-pollination of curriculum, activities, and pedagogical approaches from teacher to teacher.

Description of the Responsive Elements in Physics Teaching Communities of Practice.

In the design and enactment of the IPaSS program, three types of strategies were exercised to ensure responsivity in the CoP. Each of these is discussed further below:

1) ***Elements related to fostering a physics teaching CoP:*** In fostering a physics teaching CoP, involving teachers in the co-design and co-facilitation of PD activities was key. Having teachers present not just on how they have implemented university material in their classroom settings, but on any activities and pedagogical strategies that they have used, was suggested by a cohort 1 teacher in the spring prior to the summer PD in Year 2 of the program. We adopted the suggestion and as the teacher-led presentations were successful, we have incorporated them into the summer PD in the following years. After one year of being involved in IPaSS, the teachers were encouraged to design and present workshops on a topic of their choosing, including but not limited to their experiences of adapting university materials to their own contexts. This element created opportunities for teachers to share their expertise, experiences, and contexts. This was also aligned with the CoP model by placing teachers at the center of learning, giving them responsibility for designing and facilitating some of the PD sessions.

Additionally, starting in Year 2 of the program, intentional community-building activities and unstructured hours for reflection and bonding facilitated the development of peer relationships in the CoP. Teachers participated in optional after-hours activities such as trivia night, mini-golf, escape room, painting in a studio, dinner in a restaurant, and gathering at a local teacher's house to bring people together beyond the eight formal hours of instruction on campus. The activities were based on teacher survey responses, with one or more activities scheduled for each evening teachers were on campus.

2) ***Elements related to responsive PD:*** To ensure responsivity, we attended to teachers' individualized curricular needs and interests and tailored the program offerings to those needs. Examples include the iterative design of the PD schedule and content based on teachers' feedback and providing novice teachers with start-up materials when requested. Surveying teachers regularly about their overall satisfaction, curricular needs, pairing experiences, group discussions, our team's support, and other needs and ideas enabled us to inject changes into the program structure and schedule as the PD was happening. Additionally, teachers are given complete freedom regarding their degree of use of any instructional materials presented by peers or the university partner in the PD sessions. The program has no requirements or thresholds for teacher implementation of the materials. Teachers are trusted as the experts in their context to determine the usefulness of such materials.
3) ***Elements crossing both responsiveness and CoP:*** We believe some program structures both ensure responsivity and foster engagement

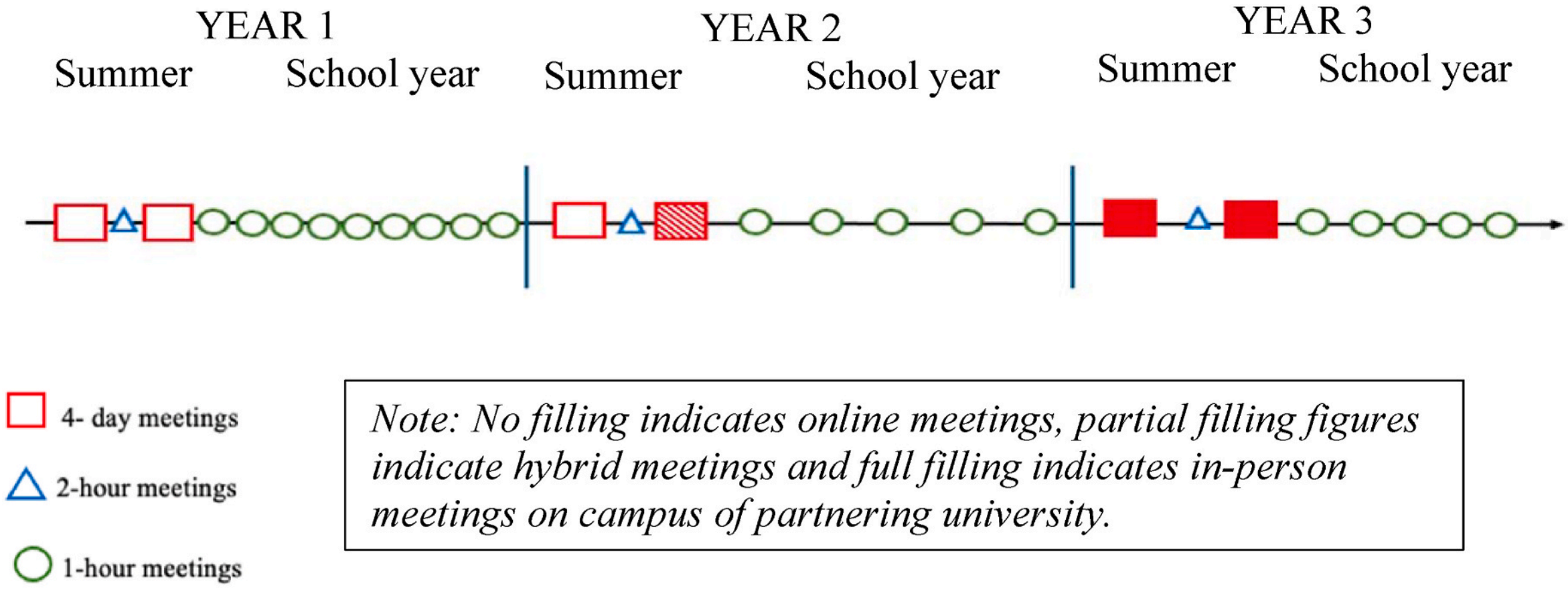

**Fig. 1.** PD Timeline Due to Covid-19, Year 1 was fully online. The frequency of online meetings in the school year changed from every week to every other week in Year 2 of the program.

in the CoP. Facilitatory responsiveness to teachers' desire to share through creating opportunities for both casual and formal share-out, and intentional but adaptive teacher grouping were among the strategies that were exercised to ensure engagement in the CoP. For instance, responsiveness to teachers' desire to share is aligned with bringing teachers to the center of learning based on the CoP framework. Similarly, the initial pairing of teachers (by PD facilitators) and subsequent adjustments (suggested by teachers) ensured pairing up of novice teachers with more experienced teachers in productive groups in an apprenticeship manner that CoP describes.

## 3. Methods

### 3.1. Focal community

Thirteen out of fourteen teachers from the first two cohorts in IPaSS (six men and seven women) participated in interviews as a part of this research project. Table 1 shows the demographic information of the teachers in IPaSS, specifying their level of experience and the physics courses they taught as of the 2021–2022 academic year. The three Cohort 1 teachers all had at least 10 years of experience and regularly taught AP Physics. The ten Cohort 2 teachers' experience and course loads were more varied, both in terms of years of experience and courses taught. In addition, five Cohort 2 teachers have non-physics backgrounds. More than 50 percent of students in participating schools (both in Cohort 1 and Cohort 2 schools) come from Title I schools.

### 3.2. Research design and data sources

In this work, we took a qualitative phenomenological approach to data collection and analysis to focus on teachers' experiences of Responsive Professional Development (RPD). In this approach, the researcher seeks the *essence* of the shared experiences of participants through in-depth interviews (Creswell, 1998, 2016; Merriam & Tisdell, 2015; Moustakas, 1994; Polkinghorne, 1989). Here, the experience or the phenomenon under investigation is online and in-person PD sessions.

Using a criterion sampling method that is appropriate for phenomenological research (Creswell, 1998), we conducted 30-min semi-structured interviews with the 13 teachers listed in Table 1. This sampling method allows researchers to gain an in-depth understanding of the participants' experiences of the phenomenon under investigation (Creswell & Poth, 2017). All interviews were conducted in Spring 2022 by the first author. The research activities were approved by the university Institutional Review Board, and all teachers consented to participate. We were unable to schedule an interview with one of the Cohort 1 teachers, so she is not included in this analysis nor in Table 1. The interviews were aimed at teachers' perceptions of the program. Teachers were given access to the interview questions at least one week before the interview through an online shared folder. One teacher provided written responses to these questions which he submitted to researchers prior to the interview, and these responses were used as part of his interview.

The interview protocol consisted of two sections relevant to this paper. In the first section, there were two questions: one question asked teachers for their general opinion of the program and the second question inquired about the extent to which this program was similar to or different from their prior PD experiences. In the second section, we inquired more specifically about their views of the program by asking both about the most helpful aspects of the program and their thoughts on improving the program. We also asked about the extent to which they identify themselves with the community of teachers.

The teachers also filled out PD surveys each day of the in-person summer PDs and before each online session. The daily surveys and observation field notes from the summer PD meetings (three years) were used for triangulation with interview results. Some examples of the daily survey questions that were asked of teachers after the completion of each day of in-person PD were *(1) How did your interactions with other teachers go? Please share at least three things that you learned from other teachers today. (please name the teachers you interacted with). (2) What was the most challenging activity for you today? Did anyone help you with resolving this challenge? (3) If you have suggestions/thoughts for the PD, please use the box below and let us know*. An example of a typical pre-PD survey item in online PD was, *Request for help: Please use this field to alert us of any targeted* support *you need from the university team. You may also use this space to provide feedback or ideas for the program more generally. For emergent issues, please give a brief description and provide your preferred timescale for addressing the issue.*

### 3.3. Data analysis procedure

The interview data was transcribed and reviewed for clarity by two researchers. All surveys and interview data were then uploaded into MAXQDA22. We employed a combination of deductive and inductive approaches for coding interviews. Pre-assigned codes were used based on the interview protocol to deductively capture the big ideas in the interviews, then, using an inductive approach, the sub-codes were created to encapsulate the variety of teachers' responses in responding to similar interview questions. The coding scheme was revised by one researcher and iterated in collaboration with two others. Together, the coders analyzed three interviews to test out and iterate the coding scheme. With the new coding scheme, all three researchers coded a fourth interview and reached an agreement of Cohen's Kappa = 0.74, resolving all disagreements through consensus discussion. The first researcher then recoded the three interviews used to iterate the coding scheme and the remaining 9 uncoded interviews. The first researcher also used the revised coding scheme to code the survey responses. The new coding scheme was further refined to merge some similar codes and create categories. A fourth coder who had not coded before recoded all interviews for the existence or non-existence of each category. The results were matched with the previous categorization of codes with a Cohen's Kappa of 0.86.

**Table 1**
Descriptive information of teacher cohorts in the second year of the IPaSS program.

| Cohort | Pseudonym | Gender | Ethnicity | Experience (years) | Courses taught |
|---|---|---|---|---|---|
| 1 (2019–2020) | Lisa | Female | White | 27 | AP Physics C[a],<br>Algebra-based Physics |
| | Jeff | Male | White | 22 | AP Physics 1[b], AP Physics C |
| | Francesca | Female | White | 14 | AP Physics 1,<br>AP Physics C,<br>Algebra-based Physics |
| 2 (2020–2021) | Kayla | Female | White | 7 | Algebra-based Physics |
| | Grant | Male | White | 17 | Algebra-based Physics, Honors Chemistry, Dual Credit Statistics, Intro to Computer programming, Earth and Space Science |
| | Tony | Male | White, Hispanic | 2 | AP Physics 1, Algebra-based Physics (including in-Spanish instruction for bilingual Spanish-speaking students) |
| | Sophia | Female | White, Hispanic | 6 | AP Physics 1, Algebra-based Physics, Astronomy, Biology |
| | Amy | Female | White | 6 | Algebra-based Physics,<br>Conceptual Physics |
| | Patrick | Male | White | 5 | AP Physics 1,<br>Astronomy, Chemistry, Biology 1, Freshman Physical Science |
| | Carl | Male | White | 30 | Algebra-based Physics,<br>Physical Science[c],<br>Chemistry |
| | Dawn[d] | Female | White | 3 | Astronomy, Biology, Chemistry, Zoology |
| | Susan | Female | White | 32 | AP Physics C,<br>Algebra-based Physics |
| | Paul | Male | White | >30 | Algebra-based Physics,<br>Physical Science,<br>General Chemistry |
| | Total = 13 | Male = 6<br>Female = 7 | | | |

[a] AP Physics C is a calculus-based college-level course.
[b] AP Physics 1 is an algebra-based college-level course.
[c] Physical science includes the introductory study of chemical, physical, earth, environmental, and life science content with emphasis on the scientific method, metric system, graphing, lab safety, technology, and career opportunities.
[d] Dawn has 2–3 years of additional experience teaching high school homeschool enrichment classes.

### 3.4. Coding scheme

The codes have been summarized under four main categories. Table 2 shows the name of each code, the category under which we summarized the codes and the description of the code.

## 4. Results

All 13 interviews included segments coded showing that teachers' experiences with the program have been positive in general. However, the characterization of positive experiences varied among participants as they highlighted different aspects of the program. The phenomenological approach allowed us to capture the nuances of their experiences and connect them to some facilitatory moves that we took in the IPaSS program. Below we report findings based on four categories that emerged when we asked teachers to talk about the most salient features of the program: 1) core design features of the program which had been initially taken into account based on the effective PD features in the literature, 2) pedagogical and instructional support from peers, 3) professional growth opportunities provided by the program, and 4) social and personal benefits of community involvement. The frequency of each of these four categories arising in the interviews is shown in Fig. 2. In the next section, we also unpack the responsive aspect of the IPaSS program that may have been involved in creating such perceptions among teachers by referring to teachers' data in each section.

1) Program Core Structures

In elaborating on the salient features of the program in the interview, teachers often compared this program with their prior PD experiences. Table 3 shows the features they discussed, including the number and percentage of segments with each code, and the number of teachers who talked about that feature. Teachers most frequently remarked on the access to instructional materials (22 segments from 11 teachers) provided by the program and sustained support from the university team (12 segments from 5 teachers). Support is often needed alongside access to instructional materials, and teachers mentioned getting support from the team as the second most important feature of the program. In their interviews, they specifically referred to the prolonged PD experience over the course of a year and the university team's support (both in-person and online) when issues arose. These five teachers were a mix of novice and veteran teachers with varied backgrounds both in physics and other sciences (i.e., the access and support codes were not exclusive to a particular group of teachers).

Another core feature of the program pointed out by teachers was the practicality of the instructional materials. This means that teachers could implement the activities in their classes with few or no modifications. For instance, a teacher with a non-physics background mentioned that he completed the summer session with "a lot of practical stuff that I could actually use in my classroom pretty much right away." Kayla, a novice teacher, also mentioned, "It's really great about the IPaSS program that I can walk away with something that I can immediately try on in my class." Practicality here was also not exclusive to a particular group of teachers, and the six teachers who talked about it were a mix of veteran, mid-career, and novice teachers. The teachers emphasized practicality over theoretical concepts, which they found a challenging feature of some other PD experiences.

**Table 2**
Coding scheme.

| Categories | Codes | Description |
|---|---|---|
| Pedagogical and Instructional Support | Exchanging and developing pedagogical *ideas* | Exchanging various ideas about the pedagogy of teaching physics |
| | Exchanging and developing instructional *materials* | Sharing both university and non-university resources entails both mutual sharing and uptake which is a one-way type of sharing. |
| Professional Growth Opportunities | Collaborative opportunities | Opportunities where the program provided opportunities for collaborative work and interaction. Also, captures specific examples when teachers talked about how in-person meetings fostered collaboration in general. |
| | Mentoring opportunities | Appreciation of provisions of mentoring opportunities. Entails both examples of sharing perceptions of seeing themselves as mentor or mentees. |
| | Moving toward leadership roles | Appreciation of opportunities to take more leadership roles and shine. |
| Social and Personal Benefits of Community Involvement | Improved confidence and belonging | Specific examples of social support such as developing confidence, sharing emotional stories, and getting support from the community. |
| | Relationship building/bonding | Examples of connecting with teachers without clear reference to pedagogical support. Also includes examples from isolated teachers. |
| Program Core Structures | Access to university materials | Access to university level materials (e.g., SmartIllinois, iOLab device) |
| | Sustained support from the university | Support from the university team (e.g., email, online PD sessions, school visits) |
| | Serving diverse teachers (e.g., diverse backgrounds/experiences/needs) | Serving teachers with both physics and non-physics degrees teaching various levels of physics |
| | Flexibility in uptake and implementation | Gradual implementation of the program materials at teachers' pace |
| | Practical materials | Ready-made labs or pre-loaded videos and homework for direct use in the class |
| | Focus on physics content knowledge | Dedicated time for teaches to adapt university-level materials |
| | In-person meetings/PD | In-person workshop meetings on campus |
| | Opportunities for reflection about teaching | Dedicated time for self-paced work which creates opportunities for reflection on teaching |
| | Responsive to teachers' needs | The general responsivity of the program to teachers' needs |
| | Bridging the gap between university and high school | Partnership with high school teachers which has created opportunities for sharing university resources with teachers |

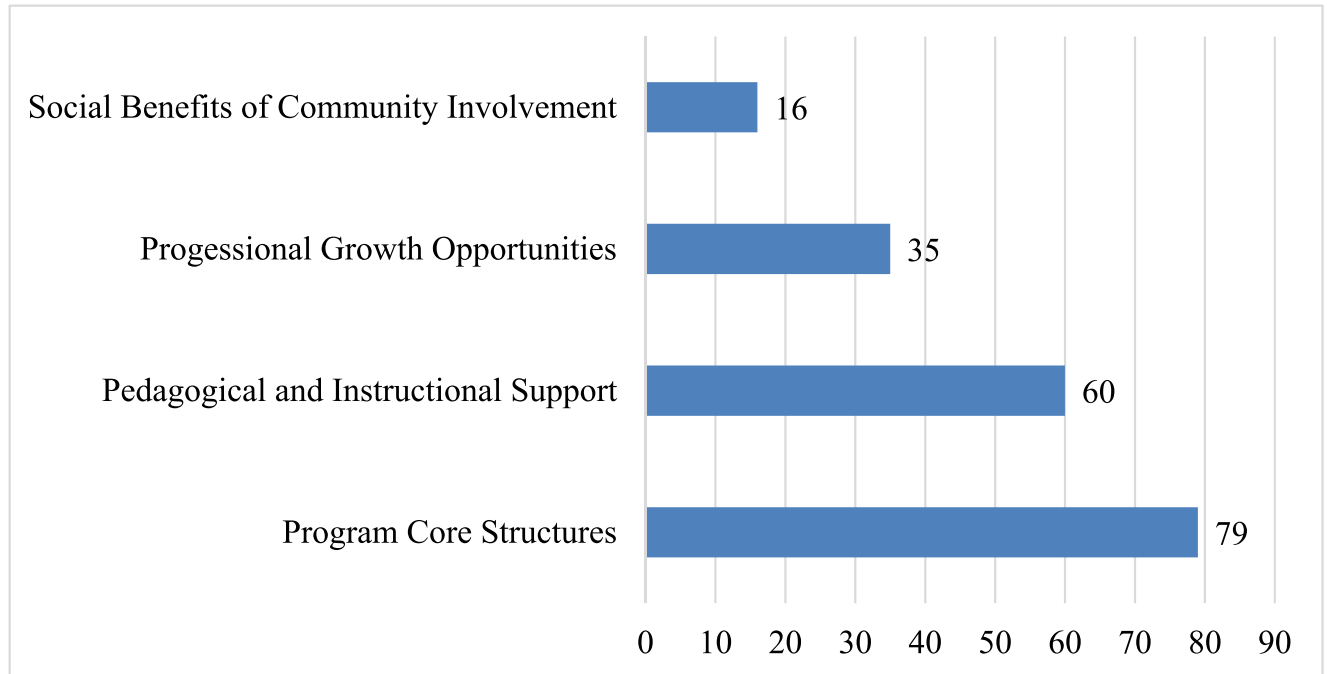

**Fig. 2.** Frequency of coded segments in each category.

**Table 3**
IPaSS Program's core structures that teachers found helpful.

| Code | Number of coded segments | Number of teachers | % coded segments |
|---|---|---|---|
| Access to university materials | 22 | 11 | 27.85 |
| Sustained support from university team | 12 | 5 | 15.19 |
| Serving diverse teachers | 9 | 6 | 11.39 |
| Flexibility in uptake and implementation | 9 | 7 | 11.39 |
| Practical materials | 9 | 6 | 11.39 |
| Focus on physics content knowledge | 5 | 3 | 6.33 |
| In-person meetings/PD | 4 | 3 | 5.06 |
| Opportunities for reflection about teaching | 4 | 3 | 5.06 |
| Responsive to teachers' needs | 3 | 3 | 3.80 |
| Bridging the gap between university and high school | 2 | 1 | 2.53 |

Flexibility in the implementation of university materials was also not limited to a specific group of teachers. Here, one of the mid-career teachers refers to flexibility in implementation, which gives them the agency to pick freely among the resources:

That big thing [that stands out] has really been the freedom and pace. You're just free to pick and choose. There's just so much [variety], so you know it's like, here are some tools, and then you can create [your own unique version of] what that looks like.

From other core features, three educators discussed in-person meetings as a notable feature that is mostly missing from other PD opportunities, especially after the COVID-19 pandemic. Despite acknowledging the challenge of holding and attending in-person meetings, they appreciated the two weeks of in-person summer workshops as an opportunity that foster collaboration among participants.

### 4.1. *Flexibility as a responsive move*

Table 3 shows that teachers identified both responsive features (e.g., flexibility in implementation) and effective PD elements from the literature (e.g., discipline-specific PD) upon which the program was based. For instance, flexibility in uptake and implementation was a responsive feature added to the program by attending to teachers' diverse backgrounds and needs. In the IPaSS, unlike many others in which our teachers were involved, we did not require them to immediately implement all university materials into their classroom practices. Hence, teachers' participation in the program was not contingent on their use of any of the university materials or approaches at all. We trusted that the teachers knew their own students and contexts and could make informed decisions about what would work best in their courses. Teachers had a chance to familiarize themselves with the materials and implement them at their own pace and to the degree that they felt was manageable for themselves and their students. For instance, in their first year participating in the program, teachers commonly try a few of our lab activities, gradually increasing the number of lab activities with the iOLab over time.

2) Pedagogical and Instructional Support from Teaching Fellows

In the second category, we have pedagogical and instructional support from peers, which is distinct from the support that the university team provides teachers. Here, the teachers appreciated both exchanging and developing pedagogical *ideas* and instructional *materials*. In studies with a phenomenological approach, it's important not to oversimplify or reduce the nuances of experiences to fully understand the subtleties of differences. This process is called horizontalization (Merriam & Tisdell, 2015). All 13 teachers had statements coded as valuing the sharing of pedagogical ideas. Specific ideas identified by teachers range from new, unique activities that they had never heard of (Amy mentioned "whiteboard speed dating") to sharing concerns (Carl talked about supporting students in AP classes). Francesca, a mid-career teacher, spoke on behalf of other singleton teachers about teacher isolation and the importance of sharing ideas for this particular population:

… coming out to [Midwest region] you know, knowing that out [here], you know there's probably [only] one physics teacher in the building, and that one physics teacher probably also teaches other things besides physics because there's just not a lot of physics [instruction] that's happening in buildings, and that is isolation and that is, you know this inhibition to get new ideas …

Teachers appreciated getting *ideas* from colleagues teaching in environments both similar to and different from their own. Teachers from different contexts found exchanging ideas with other teachers with diverse backgrounds, expertise, and school contexts very generative as "everyone is approaching things in a different manner." Teachers with similar contexts and backgrounds found that while they still appreciated "stealing ideas," they found the sharing experience to be "renewing" for them as educators.

All 13 teachers included statements about valuing *materials* from other teaching fellows, with a total of 31 coded segments. Examples of useful share-outs identified were labs (with the iOLab, non-university resources, and various instructional strategies. Notably, incoming teachers found great value in university materials that had been adapted by their colleagues in earlier cohorts who had already made adaptations to university labs: "trying to see how other teachers use it versus what I do because I can get set in my ways." In addition to adapted materials from university labs, teachers appreciated access to a myriad of resources from outside sources shared by other teaching fellows. For instance, Susan, a veteran teacher, mentioned how she learned from blog posts shared by Francesca, which significantly enriched her instruction.

We believe incorporating CoP elements such as involving teachers in co-design and co-facilitation of PD and scheduling unstructured time during in-person PD helped in creating opportunities for pedagogical and instructional support. Below we provide further evidence connecting these outcomes to PD elements.

### 4.2. *Teacher Co-design and Co-facilitation*

Giving teachers opportunities to design and run workshops brought a lot of resource-sharing that was helpful for novice, experienced, and out-of-field teachers. Francesca, identified by nine teachers as having useful resources, reflected on sharing materials through presentations and said: "That was very nice last summer, the idea of teachers doing the workshops and it was kind of just like here are my things here [are] my resources, so that's been really helpful." Similarly, Grant, a mid-career teacher with a math background, valued teachers' presentations in acquiring new ideas and pedagogical practices. Even though teachers' co-facilitation was one of the PD elements based on the CoP framework, these presentations were particularly in response to teachers' interests and requests to share ideas with the larger group.

### 4.3. Unstructured time

The semi-structured summer schedule that accommodated teachers' presentations and workshops also allowed for some unstructured time during which teachers could plan and develop curricula for the upcoming school year, discuss ideas with colleagues, seek feedback, and reflect on their teaching. Susan, a veteran teacher, explained:

You asked me to compare what was so much better about what we did over the summer [and it] was the long stretches of time where you do what you need to do, talk to [others], … so you can actually sit and have a conversation, …but having that [time and flexibility] is what I think that's what most professionals appreciate I don't understand, sometimes why educators are not treated like that.

Along the same lines, Penuel et al. (2007) found the incorporation of additional time for teachers to plan for their implementation a very useful strategy for PD.

3) Professional Growth Opportunities

One of the well-documented affordances of learning in a CoP is collaboration among people with varying levels of experience; hence, creating ample opportunities for professional growth. This can often result in the formation of mentor-mentee relationships, wherein the mentor supports the mentee's learning. Our data analysis revealed that collaboration between teachers with varying levels of experience created a vibrant CoP where mentorship is highly valued. Under the category of professional growth opportunities, mentorship opportunities were coded for in 29 segments from 12 teachers either in the form of providing mentorship or receiving it. Further, our analysis showed that the collaboration among teachers transcended the mentor-mentee dyad described in the original conceptualization of the CoP framework (Lave & Wenger, 1991). The results revealed that both novice and experienced teachers benefited from the mentorship opportunity, each in different ways. Novice teachers appreciated getting help from more experienced teachers, and veteran teachers valued novice teachers' new perspectives. For instance, Patrick, an early career teacher teaching AP physics for the first time, described his learning experience with a veteran teacher, Jeff, who has 30 years of experience:

Another good example would be the iOLab because we're given all the labs, but when I would go through them it was different from when Jeff would go through the lab with us, which was also super useful. And then from there, after going through it once, I can push it through directly to students. It's really hard to get a good PD where you can get teachers to be able to push [materials] directly to students. I appreciate when we can do that.

Breaking the typical unidirectional flow of knowledge in the CoP, veteran teachers, also, talked about gaining new perspectives by interacting with novices. Here is a quote from Lisa, a veteran teacher trying to remain open to new strategies and ideas in teaching physics as opposed to positioning herself as a veteran teacher with fixed approaches to teaching:

I'd like to think that I'm not a veteran teacher who says 'This is the way it needs to be done, this is the way I do it, this is the way you should do it' or, 'This is the way I will always do it.' Every year, I've rethought what I've done. I'm always interested in hearing how somebody approaches something. And I recall times when [other teachers] have taught me something, and I've gone, 'oh that's a new way of thinking about something' or 'that's a different way of approaching [it],' and then I've stolen [that method or tool] …

She further reflects on her gains from being a mentor in a physics teaching CoP:

… I do feel, as a veteran teacher, that I'm probably going to give more than I get, but I'm still going to get. I guess that's how I would approach it. I enjoy sharing and helping somebody develop as a teacher. I think you learn a lot from trying to help somebody solve a problem.

So, the physics teachers in IPaSS, position themselves beyond the typical mentor-mentee relationships described in the CoP framework, demonstrating that both sides benefit from learning from each other in a CoP, and that veteran and novice teachers equally value the learning they get from their interactions.

Additionally, the ways in which the IPaSS program design has helped teachers move toward leadership roles came up several times for one teacher. Francesca, who had been involved in PD co-design and facilitation for two consecutive years and was able to run her workshops for other teachers, said: "This program has given me the space to finally act on the things that I have felt very passionate about for our physics teacher community." She further expands that "I'm saying again I am so thankful to be fortunate enough to be in a position where I get to really step into a teacher leadership position in this in this way now." We believe program structures such as teacher grouping and responsivity to teachers' desire to share helped facilitate professional growth opportunities.

### 4.4. Intentional teacher groupings

Intentional grouping and responsive regrouping of teachers during in-person PD supported the formation of informal mentorships within and between teacher cohorts. New teachers in the program informally apprenticed with returning teachers during scheduled planning periods, often giving them a more tractable starting point when exploring new resources. During the program's second summer workshop, we formally paired teachers together based on their school contexts and levels of physics taught, matching up novice teachers with veteran teachers when possible. These groups completed activities together and were encouraged to meet during informal planning periods throughout the PD. Upon surveying the groups, we found that several pairs needed to be rearranged to maximize collaboration. For example, two novice teachers in the second cohort were paired together based on their teaching assignment, yet each of them commented in their daily post-PD survey that they were not able to have motivating conversations because each of them was feeling somewhat overwhelmed. Being attentive to this need, we responded with different pairings the next day, pairing these teachers with veteran teachers from the first cohort. Both second cohort teachers provided survey feedback that this new pairing was more productive. Here is a quote from Dawn, a novice teacher, after regrouping:

I feel bad that Carl is my partner every day because he's giving me so much. Then they switched up partners a little bit, so I think that was better I feel like I'm not as pulling on everybody else as we switched around.

### 4.5. Responsiveness to teachers' desire to share

In response to teachers like Patrick, who desired to share their knowledge with others, we designed an adaptive semi-structured agenda for facilitating sharing during the in-person PD. This design feature created ample opportunities for informal sharing for teachers over a shared enterprise. Teachers had a chance to interact informally with other teachers who were not necessarily more experienced than them but were teaching in the same context or level. Here is how Patrick reflects after one of these meetings:

I really liked what Tony had [developed]. It was super helpful and super easy. But with that being said, I really am looking for ways I want to create and help [others] as well. I don't want to be a full 'take' kind of addition to the group. So, I really also want to find a way to add as well. I am looking to do that in the summer. We'll see what we can do.

Patrick's readiness to take on a more active role, coupled with his passion for bringing other aspects of physics learning to the community

(e.g., transferability of data analysis skills that students learn when working with the iOLab), helped to support his smooth transition toward core membership in the group.

It is worth noting that we intentionally attended to teachers' strengths over the first year of their participation, and then met with them before their second summer to encourage them to contribute to a workshop. These conversations also had the benefit of messaging teachers that everyone – regardless of background or preparation to teach physics – has something to offer the community.

4) Social and Personal Benefits of Community Involvement

Two key benefits of community involvement emerged from our analysis of teacher responses: improving confidence and relationship building/bonding. Three teachers with varying levels of experience reflected on their experiences of improved confidence in their teaching abilities. Kayla, a novice teacher, reflected on her experience in the program and how it improved her confidence and belonging as a result of seeing other colleagues' challenges:

And so I think just over time this program allowed me to gain some confidence in my own ideas which allows me to kind of feel a little bit more like I belong with the other teachers in the program. There was the general sense of, at first, kind of almost feeling [like] I was a fraud in the program, which [I know] is all in my head. And so just over time, seeing the others [and] where they struggle on sometimes with the content or sometimes with getting students to engage with the material, things like that, and I think there's just a commonality that we've started to find over time.

Similarly, Sophia, a mid-career teacher, reflected on how her increased confidence level was linked to other teachers sharing "vulnerability":

So you … you're being there and [another member of the group] so it helps when everyone is sort of talking and sharing that you have the sense of *I can do it too*. Okay, even if I feel like I don't know if what I'm doing is right, but I can do it as well, we can share, we can. I feel that. What [do] they call it? The vulnerability. That is okay and that you can do it, and it will be [a] good group and support system that is going to allow you to do them in a way that you're going to enjoy it.

Seven teachers over 11 coded talked about relationship building and bonding. These teachers discussed the importance of interacting with each other regularly as a golden element of community building which helped them navigate some of the challenges they face in their teaching profession. For instance, Francesca, a mid-career teacher, pointed out the issue of teacher isolation (as mentioned earlier) and the social benefits singleton physics teachers can take from such communities. Similarly, Paul found "the opportunity to talk with other teachers of physics helpful" and added in parentheses: "being the lone physics teacher in the building for 30+ years."

Our results pointed to some program structures that facilitated bonding among teachers through shared challenges in a variety of ways. We believe plugging into the program elements such as community-building activities and informal conversations has helped in creating bonding opportunities. Yet, the mechanisms under which teachers experience increased confidence and belonging are not definite. Next, we discuss some of these structures.

### 4.6. *Informal conversations*

Informal conversations happen at different moments during the in-person PD as described by the following examples.

Lisa, an experienced teacher, believes there's a hidden benefit in eating meals together for bonding:

So, I like the food. Food was not [just about food], but there is something about [it] like sitting down and eating a meal with people. That changes the dynamics of what's going on. You know, just seeing what other people eat, and having those kinds of casual conversations about things. It sounds kind of goofy, but I think it really does make a difference.

Similarly, Tony, a novice teacher, appreciated informal conversations with more experienced teachers during unstructured hours. Sophia, on the other hand, attributed multiple opportunities for relationship building to the prolonged nature of the PD and its difference from one-shot PDs with a few hours of instruction:

They [other teaching experiences] are like one or two or 3 h and then they stop it. So, building on relationships? That's the big difference between IPaSS and others … It makes a huge difference when you can meet every other week with someone, to discuss something about the classroom.

### 4.7. *Designing extra-curricular activities*

Including extracurricular activities in the program created multiple opportunities for bonding and social interactions. Although there is no data recording of these after-hour interactions, teachers valued and reported having positive interactions with their colleagues. Further research needs to confirm the connection between the social and personal benefits of the community involvement category and responsiveness.

## 5. Discussion

In this work, we introduced a responsive approach to the facilitation and enactment of PD, (RPD) aiming to 1) attend to the diversity of teachers' needs, 2) adapt the PD to teachers' needs, and 3) address the inherent challenges of learning in traditional CoP framework. Theoretically, RPD is aligned with asset-based views of teacher learning in CoP, which makes teacher participation central by involving them in the process of co-design and co-facilitation of PD. We conducted this study to examine teachers' perceptions of and experiences in the RPD approach and to determine whether, in practice, responsiveness to teachers' needs achieved its intended outcomes. From 13 high school physics teacher interviews and written surveys, we identified key features of RPD that teachers identified as salient, meaningful, and beneficial. These features were aligned with our RPD design, whether based on the literature (e.g., prolonged, discipline-specific) or based on responsivity (e.g., intentional groupings). Specific instances of the core RPD facilitation approach of being attentive and adaptive were highlighted.

This research reports an exploratory, phenomenological study aiming to understand teachers' perceptions of a particular responsive PD program. We do not aim to generalize these results to all teachers in all responsive PD programs; instead, we aim to generate insights for future research, focusing on teachers' perceptions and self-reported experiences.

### 5.1. *Taking a responsive approach to teacher professional development*

Previous research on adaptive approaches to teacher PD typically evaluates the effectiveness of pre-defined adaptive strategies such as flexible implementation of materials (e.g., Penuel et al., 2011; Trautmann & MaKinster, 2010). By contrast, our work focuses not on instructional implementations, but on teachers' perceptions and experiences of the responsive features of the RPD program that contributed to their positive experiences. This approach minimizes potential bias toward specific responsive features inserted into the IPaSS by PD organizers. The teachers identified PD features tied to our core principles of responsiveness—being *attentive* and *adaptive*—that attended to multiple perspectives and empowered them to contribute to shaping their own learning experiences. Our work builds on prior approaches that prioritize leveraging teachers' assets and bringing their voices to the center of PD instruction (Koellner et al., 2011; Maskiewicz, 2015; Richards, 2022; Watkins et al., 2017, 2020). We further expanded this approach through day-by-day tailoring of the instructional practices and involving teachers both in the design and enactment of the PD sessions. As a result, teachers reported higher satisfaction with the IPaSS program in comparison to previous PD programs they had experienced. Similarly, recent research supports the benefits of placing teachers at the center of learning communities (Lin et al., 2024). This underscored the importance of empowering educators in shaping their own PD experiences.

By adopting a responsive approach in designing and implementing PD instruction, our phenomenological approach revealed that teachers in our study derived benefits such as pedagogical and instructional support, opportunities for professional growth, and personal and social benefits from community involvement. Even though such benefits were not unprecedented in prior studies (Zhang et al., 2011), having a defined set of concrete PD elements positively acknowledged by teachers as helpful was gratifying. This recognition underscores the effectiveness and relevance of responsive PD strategies in improving inequities in diverse communities, and empowering teachers (Cavendish et al., 2021).

The value of forming teaching CoP, especially in fields such as physics where isolation can be common, cannot be overstated. Sharing resources, ideas, and experiences in a responsive fashion that we describe in this work not only enriches teacher PD but can also alleviate the sense of isolation that educators might feel. The ability to discuss classroom strategies, seek advice, and receive mentorship from colleagues creates a supportive environment conducive to growth and improvement. Relationship building and bonding among educators transcends mere exchange of information; it embodies a shared journey of learning and growth, uniting individuals with diverse backgrounds and experiences towards a shared value of excellence in teaching and learning.

### 5.2. *Being responsive to address challenges in learning in communities of practice*

Overlooking veterans' learning and overfocusing on novice learning both in research and practice has been a common critique of the traditional CoP model (Eshchar-Netz & Vedder-Weiss, 2021). In this study, we were able to identify support and collaboration among teachers beyond the mentor-mentee dyad described in the traditional CoP framework (Lave & Wenger, 1991). We found that both novice and expert teachers perceived RPD elements as attentive to their needs but for different reasons. Novice teachers tended to appreciate that IPaSS allowed them to access instructional materials, meet experienced teachers from different schools, and be exposed to a variety of implementations of materials from their veteran colleagues. On the other hand, veteran teachers talked more about reflection opportunities, gaining new perspectives by working with teachers at various points in their careers (including novices), and refining their teaching skills (the latter is also reported by Fairbanks et al., 2000; Fantilli & McDougall, 2009).

Another critique directed toward the CoP model is associated with the hierarchy of power between community members with different levels of experience (Liu, 2013; Sutton & Shouse, 2019). While it is hard to completely remove the power imbalance between teachers with varying levels of expertise and experience, the program's RPD approach helped ameliorate some of these challenges. First, intentional pairings/groupings were not solely based on teachers' classroom experience levels, but other factors such as levels of physics taught and school context were taken into account. The groups were also adaptive to change upon request; hence, it created a dynamic grouping system that also allowed pairing with teachers with the same level of experience and expertise. Second, the design of IPaSS with unstructured hours allowed Cohort 1 teachers to spend some of their time individually on their areas of interest. This freedom opened up the possibility for newer teachers to learn from each other and only refer to their veteran colleagues when needed. The staggered start cohort model of IPaSS also allowed teachers returning to the program to share their expertise of the program with incoming teachers, regardless of their teaching context or background. As such, a teacher in Cohort 2 with only two years of experience could share their expertise of the iOLab with a teacher in Cohort 3 with 30 years of experience (data not included in this study) who had never seen the iOLab. Finally, the teachers also engaged in a number of social activities during the in-person summer PD outside the 8-h workday. This allowed teachers to see one another in different contexts and form bonds outside of a professional context. We believe these factors reduced the presence of a conventional power hierarchy between veterans and novices in IPaSS program.

## 6. Conclusions and Implications

This work contributes to the teacher PD literature both theoretically and practically. Theoretically, it expands the conceptualization of responsiveness from cognitively focused responses to learners' thinking to responsively supporting learning among a community of teachers. Taking a responsive approach within a CoP, the IPaSS program brought teachers to the center of learning by involving them in the co-design and co-facilitation of the PD activities. Teacher involvement notably contributed to shaping the content and structure of PD meetings, effectively addressing the challenges that arise when learners from different levels interact within the CoP. Practically, by highlighting the features of a responsive PD program through the lenses of teachers, we draw attention to approaches for responding to teachers' diverse needs and concerns in PD.

This work can help researchers of teacher education, teacher PD facilitators, and practitioners in K-12 education understand the impacts of responsive PD on teachers' PD experiences. For example, teachers can find co-designing and facilitating the PD activities a useful method to ensure content is relevant and engaging. Researchers and PD developers can focus on the transition in the CoP from peripheral participation to core members, supporting teachers in developing leadership roles within the community. Similarly, K-12 practitioners can use a responsive PD approach to ensure inclusivity in their professional learning communities (PLCs) by bringing teachers to the center of learning.

Future research on the transitions of teachers within the CoP can test the viability of the RPD approach in easing this transition for community members. Additionally, in-service teachers were the main focus of this work, so similar works can study pre-service teachers' engagement when the levels of experience and expertise are not as diverse as they were in our group. Furthermore, it is important to test these adaptive PD strategies in various settings (in the U.S. and internationally) to understand which approaches are effective across different cultural and educational contexts.

## Funding

This research was funded by NSF DRK-12 grant number 2010188.

## CRediT authorship contribution statement

**Hamideh Talafian:** Writing – review & editing, Writing – original draft, Project administration, Methodology, Formal analysis, Data curation, Conceptualization. **Morten Lundsgaard:** Writing – review & editing, Writing – original draft, Visualization, Project administration, Investigation, Funding acquisition, Conceptualization. **Maggie Mahmood:** Writing – review & editing, Writing – original draft, Project administration, Investigation, Funding acquisition, Data curation. **Devyn Shafer:** Writing – review & editing, Writing – original draft, Formal analysis. **Tim Stelzer:** Writing – review & editing, Supervision, Investigation, Funding acquisition, Formal analysis, Conceptualization. **Eric Kuo:** Writing – review & editing, Supervision, Investigation, Funding acquisition.

## Declaration of competing interest

The authors declare that they have no known competing financial interests or personal relationships that could have appeared to influence the work reported in this paper.

## Acknowledgments

The authors would like to thank the teachers of the professional development program for participating in this research.

## Appendix A. Supplementary data

Supplementary data to this article can be found online at https://doi.org/10.1016/j.tate.2024.104812.

## Data availability

Data will be made available on request.